\documentclass[english,3p]{elsarticle}
\usepackage[charter]{mathdesign}

\usepackage[T1]{fontenc}
\usepackage[utf8]{inputenc}
\usepackage{rotfloat}
\usepackage{amsmath}
\usepackage{stmaryrd}
\usepackage{graphicx}

\makeatletter

\providecommand{\tabularnewline}{\\}

\usepackage{amsthm}
\journal{arXiv}

\@ifundefined{showcaptionsetup}{}{%
 \PassOptionsToPackage{caption=false}{subfig}}
\usepackage{subfig}
\makeatother

\usepackage{babel}
\begin{document}

\begin{frontmatter}{}

\title{Pricing Options with Exponential L\'evy Neural Network}

\author[labela]{Jeonggyu Huh\corref{cor1}}

\address[labela]{School of Computational Sciences, Korea Institute for Advanced Study,
Seoul 02455, Republic of Korea}

\cortext[cor1]{corresponding author. Tel: 82-10-8960-0122\\
 E-mail address: aifina2018@kias.re.kr}
\begin{abstract}
In this paper, we propose the exponential L\'evy neural network (ELNN)
for option pricing, which is a new non-parametric exponential L\'evy
model using artificial neural networks (ANN). The ELNN fully integrates
the ANNs with the exponential L\'evy model, a conventional pricing
model. So, the ELNN can improve ANN-based models to avoid several
essential issues such as unacceptable outcomes and inconsistent pricing
of over-the-counter products. Moreover, the ELNN is the first applicable
non-parametric exponential L\'evy model by virtue of outstanding
researches on optimization in the field of ANN. The existing non-parametric
models are rather not robust for application in practice. The empirical
tests with S\&P 500 option prices show that the ELNN outperforms two
parametric models, the Merton and Kou models, in terms of fitting
performance and stability of estimates.
\end{abstract}

\end{frontmatter}{}

\section{Introduction}

A L\'evy process is a stochastic process with independent and stationary
increments, roughly speaking, which is a jump-diffusion process generalized
so that its sample paths allow an infinite number of jumps on a finite
time interval. For a long time, it has been widely employed for option
pricing to overcome limitations of the Black-Scholes model \citep{black1973pricing},
which is inconsistent with several well-known facts such as volatility
skews, heavy tails of return distributions and so on (cf. \citet{hull2016options}).
The weakness of the model comes from that it only uses a Gaussian
process for modeling market returns. However, it is possible for the
returns to follow various distributions by using a L\'evy process
which is to add a jump process to the Gaussian process. In this context,
a variety of researchers have proposed many models based the L\'evy
processes, the majority of which fall under a category of the exponential
L\'evy (exp-L\'evy) models. The exp-L\'evy models are broadly divided
into two different types: parametric exp-L\'evy models (\citet{merton1976option,kou2002jump,madan1998variance,carr2003stochastic})
and non-parametric exp-L\'evy models (\citet{cont2004nonparametric,belomestny2006spectral}).
In general, compared to the parametric models, the non-parametric
models have great potentials to give better fitting results due to
a large number of parameters. However, to calibrate the non-parametric
models well, one should find a ``good'' minimum, almost as good
as the global minimum, of a very high-dimensional objective function,
but it is extremely challenging because the function has a bumpy surface.
So, the non-parametric models often need techniques to stabilize the
objective functions, such as regularizations and Bayesian priors.
In literature, \citet{cont2004nonparametric} penalize an objective
function by the relative entropy with respect to a prior, and \citet{belomestny2006spectral}
choose a cutoff value of spectral domain and exclude the information
outside the value. Unfortunately, even the techniques for the non-parametric
models were unsuccessful in achieving the good minimum.

The works on artificial neural networks (ANN), which have been rapidly
evolving after the outstanding success of \citet{hinton2006fast},
can give more desirable solutions to the optimization problem. The
ANN emulates a complex structure of a biological brain, so it can
learn tasks by considering numerous examples. According to the universal
approximation theorem, the ANN is able to approximate continuous functions
on compact sets under mild conditions (refer to \citet{cybenko1989approximation,hornik1991approximation}).
In addition, it does not require feature engineering that an experiment
designer manually extracts features from given data, thereby helping
to discover hidden principles producing observations without any biased
view. Notwithstanding these advantages, it is demanding for the ANN
to make full use of its ability because its complex structure makes
it difficult to find a good minimum. As a result, many studies has
continued to apply prevalent optimization methods to the ANN such
as the stochastic gradient method and its extensions (\citet{bousquet2008tradeoffs,sutskever2013importance,kingma2014adam}).
In this paper, we invent a new non-parametric exp-L\'evy model and
calibrate it satisfactorily by deploying the ANN and the relevant
techniques. We call the model of this paper the exp-L\'evy neural
network (ELNN).

On the other hand, the ANN has been extensively utilized for option
pricing. Most early works (\citet{malliaris1993neural,hutchinson1994nonparametric,yao2000option})
exploited the ANN as just a tool for nonlinear regression. Several
authors of the works used to claim that their approaches were attractive
because they do not require any economic assumptions. However, the
approaches without such assumptions may yield unfavorable results
in three respects. First, when it comes to a product to which small
amount of data are given (i.e., deep out-of-the-money options), the
predictions of the networks can be quite incorrect (\citet{bennell2004black}).
Second, the networks can produce unacceptable outcomes such as discontinuous
prices, which is a more serious problem than the foregoing. Worst
of all, arbitrage opportunities can occur within the estimated prices
(\citet{lajbcygier2004improving,yang2017gated}). Finally, the networks
are valid only for target options used as learning data, so they can
not give any meaningful results for the other options, so to speak,
over-the-counter products. To cope with the difficulties, the methods
to integrate the ANNs with conventional pricing models, named as hybrid
models, have been variously devised. They can be classified by the
level of the integration: weak hybrid models (\citet{lajbcygier1997improved,andreou2008pricing,wang2009nonlinear})
and strong hybrid models (\citet{luo2017neural,yang2017gated}). Under
a weak hybrid model, the conventional models and the ANNs complement
to each other in a relatively simple way. For example, \citet{lajbcygier1997improved}
corrected the model prices of Black and Scholes with an ANN, and \citet{wang2009nonlinear}
improved pricing performance of an ANN by using the GARCH volatilities
as its inputs. In contrast, a strong hybrid model is to fully combine
the ANNs with the conventional approaches. Recently, \citet{luo2017neural}
and \citet{yang2017gated} put ANNs into an one-factor stochastic
volatility model and a local volatility model, respectively, in which
the ANNs are inseparable from the conventional models. Both the strong
and weak hybrid models can rectify the wrong predictions due to sparse
data. However, only the strong hybrid models can fairly reduce unacceptable
events caused by absence of economic assumptions and be naturally
extended for pricing and hedging of over-the-counter products.

The ELNN belongs to the class of the strong hybrid models so as to
avoid the essential issues of the non-hybrid models and the weak hybrid
models.  Considering that \citet{luo2017neural} addressed stock
predictions (not option prices), \citet{yang2017gated} made only
the strong hybrid model concerning option pricing up to now. But the
model can hardly be from the limitations of local volatility models.
Particularly, the local volatility models give inadequate prices to
path-dependent options because they can not deal with conditional
events theoretically (cf. \citet{wilmott2007paul}). L\'evy's framework,
including the ELNN, is based upon advanced probability theory, so
it can provide more desirable solutions to the complex products. In
addition, the ELNN can outperform the existing exp-L\'evy models
because it receives benefits of outstanding researches on finding
a good minimum in the field of ANN. We experimentally verify it with
option prices on S\&P 500 for 5 years by showing that the ELNN fits
the data better and its estimates are more stable than two parametric
exp-L\'evy models: the Merton and Kou models. The performance of
the ELNN pans out well because it can accurately estimate L\'evy
densities even under various noises unlike the other non-parametric
models. We also prove it under virtual markets generated with the
two parametric models. It is encouraging in that the existing non-parametric
exp-L\'evy models can not be superior to several parametric exp-L\'evy
models in spite of their large number of parameters. On the other
hand, option prices should be transformed using the Fourier transform
for a training of the ELNN. However, daily data in an actual market
is generally too illiquid to be transformed precisely. We resolve
this problem by inventing a technique ``data amplification''. This
is also quite an important contribution of this paper. 

The paper is structured as follows. The next section reviews the exp-L\'evy
model and a pricing method for the model. In Section \ref{sec:ELNN},
the ELNN is introduced and its structure is explained in detail. In
Section \ref{sec:Numerical-tests}, we implement and test the ELNN
under virtual markets. Moreover, stability tests are progressed on
the existing non-parametric exp-L\'evy models. Section \ref{sec:Empirical-tests}
provides the results on empirical tests with market data. Section
\ref{sec:Conclusion} concludes.

\section{\label{sec:exp_levy_models}Exp-L\'evy models}

\subsection{Exp-L\'evy models}

As said in the introduction, L\'evy processes are often used to depict
dynamics of market returns, which can be roughly considered as a jump-diffusion
process allowing for an infinite number of jumps on a finite time
interval. The L\'evy-It\^o decomposition clarifies this perspective:
any L\'evy process $\left(X_{t}\right)_{t\geq0}$ can be decomposed
into the sum of a Gaussian process with a linear drift and two pure
jump processes, which are associated with big jumps and small jumps,
respectively. To be concrete, there exists the L\'evy-Khinchine triplet
$\left(\sigma,\nu,b\right)$ of $\left(X_{t}\right)_{t\geq0}$ such
that
\[
X_{t}=bt+\sigma W_{t}+X_{t}^{l}+\lim_{\epsilon\searrow0}X_{t}^{\epsilon},
\]
where
\[
X_{t}^{l}=\int_{\left|x\right|\geq1,s\in\left[0,t\right]}xJ^{X}\left(ds\times dx\right),\quad X_{t}^{\epsilon}=\int_{\epsilon\leq\left|x\right|<1,s\in\left[0,t\right]}x\left\{ J^{X}\left(ds\times dx\right)-\nu\left(dx\right)ds\right\} .
\]
Here, $W_{t}$ is the Brownian motion, $J^{X}$ is the jump measure
of $\left(X_{t}\right)_{t\geq0}$ which is a Poisson random measure
with an intensity measure $\nu\left(dx\right)dt$, and $\nu$ is a
$\sigma$-finite measure on $\mathbb{R}\backslash\left\{ 0\right\} $,
called the L\'evy measure of $\left(X_{t}\right)_{t\geq0}$, verifying
$\int_{\mathbb{R}\backslash\left\{ 0\right\} }\left(1\land x^{2}\right)\nu\left(dx\right)<\infty$.
Without the jump processes in the above expression for $\left(X_{t}\right)_{t\geq0}$,
it would be no different from a return process for the Black-Scholes
model.

The distribution of $\left(X_{t}\right)_{t\geq0}$ is expressed by
its characteristic function
\begin{equation}
\Phi_{X_{t}}\left(w\right)=E\left[e^{iwX_{t}}\right]=\exp\left(t\left(-\frac{1}{2}\sigma^{2}w^{2}+ibw+f\left(w\right)\right)\right),\label{eq:Levy-Khinchine}
\end{equation}
where
\begin{align*}
f\left(w\right)= & \int_{-\infty}^{\infty}\left(e^{iwx}-1-iwx{\rm 1}_{\left|x\right|\leq1}\right)\nu\left(dx\right).
\end{align*}
This expression is known as the L\'evy-Khinchine formula.

Given the L\'evy process $\left(X_{t}\right)_{t\geq0}$, an asset
price process $\left(S_{t}\right)_{t\geq0}$ is modeled as follows:
\[
S_{t}=S_{0}\exp\left(rt+X_{t}\right),
\]
under a risk-neutral measure $Q$, where $r$ is the risk-free rate
assumed as a constant. This type of model is called the exponential
L\'evy (exp-L\'evy) model. In financial applications, it is usually
assumed for feasible computations that there exists the L\'evy density
$\frac{d\nu}{dx}$ of $\left(X_{t}\right)_{t\geq0}$. Moreover, in
order for $\left(e^{-rt}S_{t}\right)_{t\geq0}$ to be a well-defined
martingale process, the triplet $\left(\sigma,\nu,b\right)$ of $\left(X_{t}\right)_{t\geq0}$
needs to meet the following conditions (cf. \citet{tankov2003financial})
\begin{description}
\item [{{[}a1{]}}] $\int_{\left|x\right|\geq1}e^{x}\nu\left(dx\right)<\infty,$
\item [{{[}a2{]}}] $b=-\frac{1}{2}\sigma^{2}-\int_{-\infty}^{\infty}\left(e^{x}-1-x{\rm 1}_{\left|x\right|\leq1}\right)\nu\left(dx\right)=-\frac{1}{2}\sigma^{2}-f\left(-i\right).$
\begin{figure}[t]
\centering{}\includegraphics[scale=0.62]{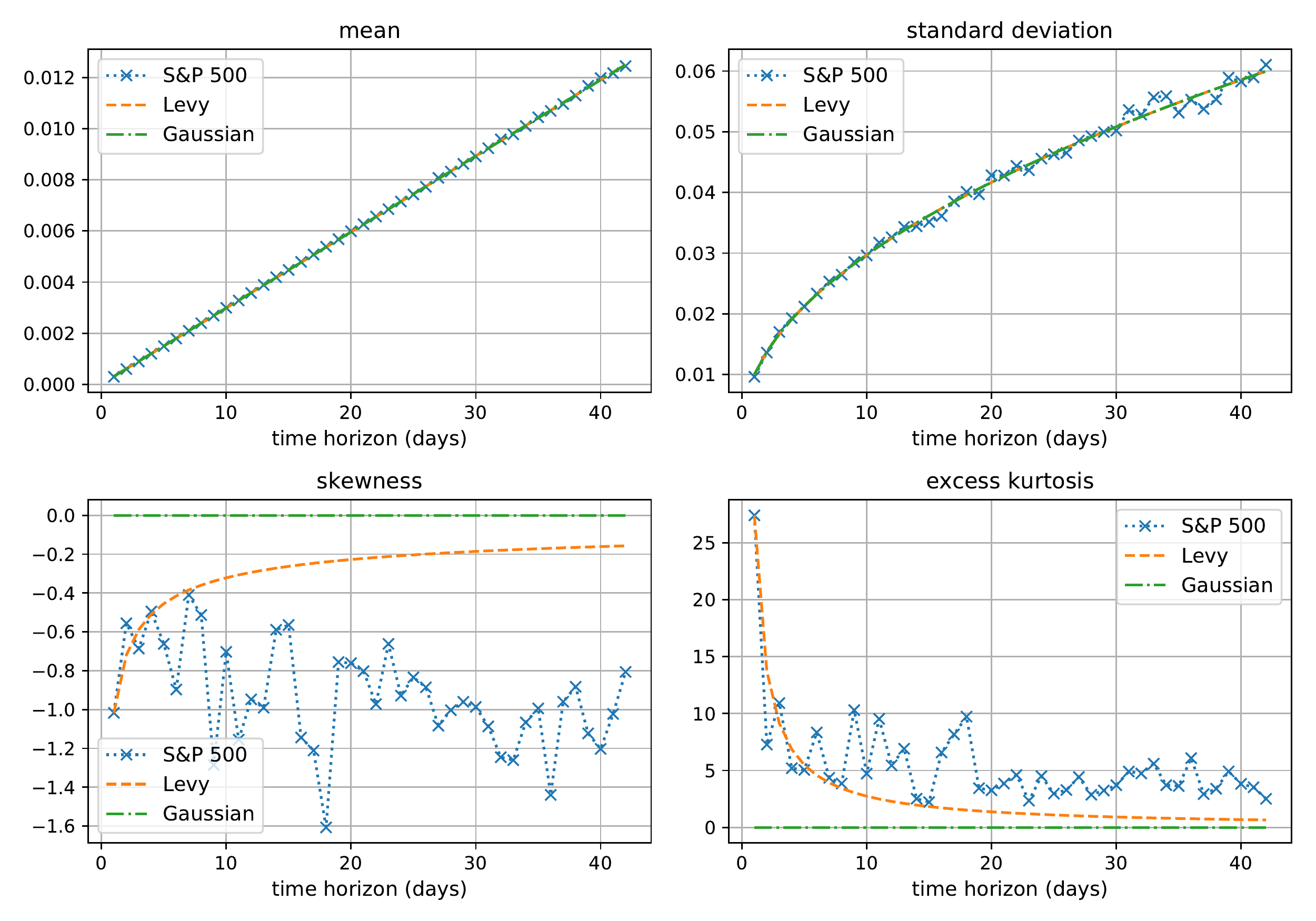}\caption{\label{fig:moments}Drawn using the return sets $R\left(\Delta\right)$
calculated from closing prices for S\&P 500, this figure shows several
moments of $R\left(\Delta\right)$ against time horizons $\Delta$.
In addition, theoretically predicted values of the moments are indicated
for the cases that the daily returns of S\&P 500 follow a L\'evy
process or a Gaussian process.}
\end{figure}
\end{description}
Under the exp-L\'evy model, the L\'evy process $X_{t}$ represents
the returns on the asset $S_{t}$ excluding interest charges. One
can notice the relation through the following equation:
\[
X_{t+\Delta}-X_{t}=\log\left(\frac{S_{t+\Delta}}{S_{t}}\right)-r\Delta
\]
for a time horizon $\Delta$. L\'evy processes are capable of reflecting
asymmetry and leptokurtosis of unconditional distribution of market
returns at a fixed short time horizon (not for wide range of time
horizons). Let us define $R\left(\Delta\right):=\left\{ X_{t_{i}+\Delta}-X_{t_{i}}\right\} $
for $t_{i}=t_{0}+i\Delta$ and $i=0,\cdots,N_{\Delta}$. Based on
$R\left(\Delta\right)$ calculated from closing prices for S\&P 500,
Figure \ref{fig:moments} shows several moments of $R\left(\Delta\right)$
against time horizons $\Delta$. The data period ranges from 3rd January,
1950 to 19th January, 2018. In addition, we indicate theoretically
predicted values of the moments for the cases that the daily returns
of S\&P 500 follow a L\'evy process or a Gaussian process. In the
L\'evy case, the mean and the standard deviation should increase
proportionally to $\Delta$ and $\Delta^{1/2}$, respectively, and
the skewness and the kurtosis should decay as $\Delta^{-1/2}$ and
$\Delta^{-1}$, respectively. On the other hand, in the Gaussian case,
the mean and the standard deviation should increase proportionally
to $\Delta$ and $\Delta^{1/2}$, respectively, as in the L\'evy
case, and the skewness and the excess kurtosis should be zeros. In
the figure, the predicted values are quite exact in respect of mean
and standard deviation, but they are far different from the actual
moments with regard to skewness and kurtosis. The Gaussian case gives
totally unreasonable values for skewness and kurtosis. The values
on the L\'evy assumption better fit the actual data, but they seem
to fall overquickly. This test demonstrates the abilities and limitations
of the exp-L\'evy models well. In fact, this conclusion is well-known
in literature. Many papers proposed two-scale models to explain fast
decaying moments at fine scales $\left(\Delta\ll1\right)$ and slowly
decaying moments at coarse scales $\left(\Delta\gg1\right)$ (cf.
\citet{bakshi1997empirical,bates1996jumps,bates2000post}). We of
course do not assert that the exp-L\'evy models agree with actual
markets; the reason for introducing the models is that they can be
considered as cornerstones to extend to the two-scale models.

\subsection{Well-known examples of exp-L\'evy models}

There are many models which belong to the category of the exp-L\'evy
models. But, among them, the most well-known are, in our opinion,
the three parametric models (\citet{merton1976option,kou2002jump,carr2003stochastic})
and the two non-parametric models (\citet{cont2004nonparametric,belomestny2006spectral}).
The variance gamma model \citep{madan1998variance} is often mentioned
but we will not go over it because the CGMY model below includes the
model. We give brief explanations for the models along with their
L\'evy densities $\frac{d\nu}{dx}$ as listed below. 
\begin{enumerate}
\item the Merton model\\
\begin{equation}
\frac{\lambda}{\delta\sqrt{2\pi}}\exp\left\{ -\frac{\left(x-\mu\right)^{2}}{2\delta^{2}}\right\} \label{eq:merton_density}
\end{equation}
Under this model, jumps occur with the frequency $\lambda$ on average,
and their sizes follow a normal distribution $N\left(\mu,\delta^{2}\right)$.
This model is simple to understand but not satisfactory to capture
the behaviors of actual jumps. It generates return distributions of
which tails are heavier than Gaussian, but all the moments of $\left(S_{t}\right)_{t\geq0}$
for this model are finite. Unfortunately, the higher-order moments
of actual assets do not seem to be finite.
\item the Kou model\\
\begin{equation}
p\lambda\lambda_{+}e^{-\lambda_{+}x}{\rm 1}_{x>0}+\left(1-p\right)\lambda\lambda_{-}e^{-\lambda_{-}\left|x\right|}{\rm 1}_{x<0}\label{eq:kou_density}
\end{equation}
Its sample path jumps with the upward frequency $p\lambda$ and the
downward frequency $\left(1-p\right)\lambda$ on average. The sizes
of the upward jumps and the downward jumps follow two distinct exponential
distributions, $\exp\left(\lambda_{+}\right)$ and $\exp\left(\lambda_{-}\right)$,
respectively. This model better reflects strong asymmetry and excess
kurtosis of the asset distribution than the Merton model.
\item the Carr, Geman, Madan, and Yor (CGMY) model\\
\[
\frac{\lambda}{\left|x\right|^{1+\alpha}}\left(e^{-\lambda_{+}x}{\rm 1}_{x>0}+e^{-\lambda_{-}\left|x\right|}{\rm 1}_{x<0}\right)
\]
This model is similar to Kou's model but does not have any diffusion
components, i.e., $\sigma=0$. Instead, it produces infinitely many
small jumps, which is a good substitute for a diffusion process in
that the small jumps also contribute to construct a leptokurtic distribution.
Meanwhile, $\alpha$ and $\lambda$ control the distribution of the
small jumps and the frequency of the whole jumps, respectively.
\item the Cont and Tankov (CT) model 
\[
\sum_{i=1}^{N}\nu_{i}\delta\left(x-x_{i}\right),
\]
where $x_{i}=x_{0}+i\Delta x$ for $i=0,1,\cdots,N$, and $\delta$
is the Dirac delta function. This model only considers finite L\'evy
measures, so it is assumed that $\sum_{i=1}^{N}\left|\nu_{i}\right|<\infty$.
Because its high degrees of freedom makes estimating $\nu_{i}$ very
unstable, Cont and Tankov penalize an objective function by the relative
entropy with respect to a Bayesian prior, by which their calibrations
considerably rely on a choice of the prior.
\item the Belomestny and Rei\ss$\,$ (BR) method
\[
e^{-x}\mathcal{F}^{-1}\left[\frac{1}{T}\log\left(1+e^{-iwrT}iw\left(1+iw\right)\mathcal{F}\left[z_{X_{T}}\left(k\right)\right]\right)+\left(\frac{\hat{\sigma}^{2}}{2}w^{2}-\frac{\hat{\sigma}^{2}}{2}-\hat{b}+\hat{\lambda}\right)-i\left(\hat{\sigma}^{2}+\hat{b}\right)w\right]\left(x\right),
\]
where $\mathcal{F}$ is the Fourier transform operator $h\left(x\right)\rightarrow\hat{h}\left(w\right)$,
$\mathcal{F}^{-1}$ is its inverse, $\lambda=\int_{-\infty}^{\infty}\nu\left(dx\right)$,
$k=\log K/S_{0}$, $K$ and $T$ are the strike and the time to maturity
of an option, respectively, and $z_{X_{T}}$ is a function involved
in the time value of the option. In fact, this is not an ordinary
model but a calibration method to estimate the triplet $\left(\sigma,\nu,b\right)$
of a market generating process $\left(X_{t}\right)_{t\geq0}$. As
the CT model does, this method also concerns only finite L\'evy measures.
It first produces an estimate $\left(\hat{\sigma},\hat{b},\hat{\lambda}\right)$
of $\left(\sigma,b,\lambda\right)$ and subsequently calculates $\nu$
by putting the estimate into the above expression. This approach is
theoretically appealing but is not applicable in practice because
it is severely ill-posed. 
\end{enumerate}
The stated models are arranged in the order of historical development.
For the parametric models, note that the parameter $\lambda_{+}$
should be restricted in a domain where their L\'evy measures obey
the assumption $\left[{\rm a1}\right]$. On the other hand, for the
non-parametric models, it is expected that $\left[{\rm a1}\right]$
is implicitly satisfied in a calibration process.

\subsection{Pricing options using the Fourier transform}

In this subsection, we review the option pricing method of \citet{carr1999option},
for which the assumption $\left[{\rm a1}\right]$ is replaced with
the following one :
\begin{description}
\item [{{[}a1{*}{]}}] $\exists\alpha>0,\;\int_{\left|x\right|\geq1}e^{\left(1+\alpha\right)x}\nu\left(dx\right)<\infty.$
\end{description}
It is a slight extension of $\left[{\rm a1}\right]$ in that $\alpha$
can be chosen as an arbitrarily small number. Before moving on, we
specify the relation between a function $h\left(x\right)$ and its
Fourier transform $\hat{h}\left(w\right)$ as follows:
\[
\hat{h}\left(w\right)=\int_{-\infty}^{\infty}h\left(x\right)e^{ixw}dx,\quad h\left(x\right)=\frac{1}{2\pi}\int_{-\infty}^{\infty}\hat{h}\left(w\right)e^{-ixw}dw.
\]
This specification is necessary because various definitions exist
for the Fourier transform. Moreover, we denote the Fourier transform
operator and its inverse as $\mathcal{F}$ and $\mathcal{F}^{-1}$,
respectively. So, $\mathcal{F}\left(h\right)=\hat{h}$ and $\mathcal{F}^{-1}\left(\hat{h}\right)=h$.

We will calculate an European call price $c_{X}$ with strike $K$
and maturity $T$ at $t=0$. From the martingale pricing approach,
it is given by 
\[
c_{X}\left(S_{0},K;T\right)=e^{-rT}E_{0}^{Q}\left[\left(S_{T}-K\right)^{+}\right].
\]
Letting $k=\log\frac{K}{S_{0}}$,

\[
\tilde{c}_{X}\left(k;T\right):=\frac{c_{X}\left(S_{0},K;T\right)}{S_{0}}=e^{-rT}E_{0}^{Q}\left[\left(e^{rT+X_{T}}-e^{k}\right)^{+}\right].
\]
By using the Fourier transform under the assumptions $\left[{\rm a1^{*}}\right]$
and $\left[{\rm a2}\right]$, $\tilde{c}_{X}$ can be achieved as
follows:
\begin{equation}
\tilde{c}_{X}\left(k;T\right)=z_{X_{T}}\left(k\right)+\left(1-e^{k-rT}\right)^{+},\label{eq:pricing_formula}
\end{equation}
where
\begin{equation}
z_{X_{T}}\left(k\right)=\mathcal{F}^{-1}\left[\zeta_{X_{T}}\left(w\right)\right]\left(k\right),\label{eq:z}
\end{equation}
\begin{equation}
\zeta_{X_{T}}\left(w\right)=e^{iwrT}\frac{\Phi_{X_{T}}\left(w-i\right)-1}{iw\left(1+iw\right)}.\label{eq:zeta}
\end{equation}
Here, $\Phi_{X_{T}}$ is the characteristic function of $X_{T}$.
Note that $\Phi_{X_{T}}\left(w-i\right)$ is well-defined and analytic
for $w\in\mathbb{R}$ by $\left[{\rm a1}^{*}\right]$, and $\Phi_{X_{T}}\left(-i\right)=1$
by $\left[{\rm a2}\right]$. These facts imply that $\zeta_{X_{T}}$$\left(0\right)$
is finite. Thus, noting that $\Phi_{X_{T}}\left(w-i\right)$ is bounded
for $w\in\mathbb{R}$, we can conclude $\zeta_{X_{T}}\in L^{p}\left(\mathbb{R}\right)$
for $p\in\left(0.5,\infty\right]$. Moreover, the real and imaginary
parts of $\zeta_{X_{T}}$ are respectively even and odd because $\zeta_{X_{T}}$
is the Fourier transform of the real function $z_{X_{T}}$.

This method is very effective in terms of calibration. First, one
can think option data over a long period of time as one-day big data
because the pricing formula (\ref{eq:pricing_formula}) does not require
the spot price $S_{0}$. What's more, for each $T$, $\tilde{c}_{X}\left(k;T\right)$
on a large range of $k$ can be quickly computed at a time by employing
the fast Fourier transform (FFT). In summary, it is achievable to
compute the model prices for numerous options with just several FFTs.
Refer to \citep{heath2002scientific,tankov2003financial} for a detailed
explanation concerning numerical implementations of FFT.

Before closing this section, we discuss $\Phi_{X_{T}}\left(w-i\right)$,
which plays an important role for a training of a network introduced
later. Denoting the density function of $X_{T}$ by $\rho_{X_{T}}$,
$\Phi_{X_{T}}\left(w-i\right)$ is the Fourier transform of $e^{x}\rho_{X_{T}}$.
Thus, by Plancherel's theorem (cf. \citet{folland2013real}), the
following relation can be obtained:
\begin{align}
\int_{-\infty}^{\infty}\left(\Phi_{X_{T}}\left(w-i\right)-\Phi_{X_{T}}^{*}\left(w-i\right)\right)^{2}dw & =\int_{-\infty}^{\infty}\left(e^{x}\rho_{X_{T}}\left(x\right)-e^{x}\rho_{X_{T}}^{*}\left(x\right)\right)^{2}dx,\label{eq:planchrel}
\end{align}
where $\rho_{X_{T}}^{*}$ is the density of another L\'evy process
$X_{t}^{*}$. This implies that, in the $L^{2}$ sense, finding $\Phi_{X_{T}}^{*}\left(w-i\right)$
closest to $\Phi_{X_{T}}\left(w-i\right)$ leads to finding $e^{x}\rho_{X_{T}}^{*}\left(x\right)$
closest to $e^{x}\rho_{X_{T}}\left(x\right)$.

\section{\label{sec:ELNN}Exp-L\'evy Neural network}

In this section, we find the best substitute $\left(X_{t}\right)_{t\geq0}$
of the market generating process $(X_{t}^{*})_{t\geq0}$. They are
assumed to be L\'evy processes, and their triplets are $\left(\sigma,\nu,b\right)$
and $\left(\sigma^{*},\nu^{*},b^{*}\right)$, respectively. In addition,
we suppose the following conditions along with $\left[{\rm a2}\right]$
given in Section \ref{sec:exp_levy_models}:
\begin{description}
\item [{{[}a1{*}{*}{]}}] $\int_{-\infty}^{\infty}e^{2x}\nu\left(dx\right)<\infty,$
\item [{{[}a3{]}}] $\int_{-\infty}^{\infty}\nu\left(dx\right)<\infty.$
\end{description}
The condition $\left[{\rm a1}^{**}\right]$ means that the variance
of the asset process $\left(S_{t}\right)_{t\geq0}$ exists, which
is generally acceptable for financial data. We set $\left[{\rm a3}\right]$
so that only finite L\'evy measures are considered in this paper.
The BR method, the other non-parametric approach, also requires $\left[{\rm a1}^{**}\right]$,
$\left[{\rm a2}\right]$ and $\left[{\rm a3}\right]$. (Meanwhile,
the CT model needs $\left[{\rm a1}^{*}\right]$, $\left[{\rm a2}\right]$
and $\left[{\rm a3}\right]$.) Under these conditions, assuming that
$\frac{d\nu}{dx}$ exists, $e^{x}\frac{d\nu}{dx}\in L^{1}\left(\mathbb{R}\right)\cap L^{2}\left(\mathbb{R}\right)$
can be derived by a straightforward computation using the Cauchy–Schwarz
inequality.

First, we have to determine what data is used for network learning.
A naive approach is to minimize frequently used measures of the difference
between the market prices $\tilde{c}_{X}^{*}$ and the values $\tilde{c}_{X}$
predicted by a network. However, to implement it, the FFT should be
performed at each epoch, which is ineffective because stable learning
needs a great amount of iterations. Alternatively, after reflecting
on the best applicable measure, we decide to minimize the $L^{2}$
distance between $\text{\ensuremath{\Phi}}_{X_{T}}\left(w-i\right)$
and $\text{\ensuremath{\Phi}}_{X_{T}}^{*}\left(w-i\right)$ for learning
of our network. By the relation (\ref{eq:planchrel}), this work may
give $e^{x}\rho_{X_{T}}\left(x\right)-e^{x}\rho_{X_{T}}^{*}\left(x\right)<\epsilon\left(x\right)$
for small $\epsilon\left(x\right)$, that is,
\[
\rho_{X_{T}}\left(x\right)-\rho_{X_{T}}^{*}\left(x\right)<e^{-x}\epsilon\left(x\right).
\]
Our method has a risk that $\rho_{X_{T}}\left(x\right)$ may differ
quite a bit from $\rho_{X_{T}}^{*}\left(x\right)$ for $x\ll0$. On
the other hand, $\Phi_{X_{T}}^{*}\left(w-i\right)$ can be computed
in the following way:
\begin{equation}
\Phi_{X_{T}}^{*}\left(w-i\right)=1+e^{-iwrT}iw\left(1+iw\right)\mathcal{F}\left[z_{X_{T}}^{*}\left(k\right)\right]\left(w\right).\label{eq:phi*}
\end{equation}
This expression is easily deduced from the two formulas (\ref{eq:z})
and (\ref{eq:zeta}).

Now, we try to express $\Phi_{X_{T}}\left(w-i\right)$ explicitly.
By the L\'evy-Khinchine formula (\ref{eq:Levy-Khinchine}) and the
assumption $\left[{\rm a2}\right]$,
\begin{align}
\Phi_{X_{T}}\left(w-i\right) & =\exp\left(T\left(\left(-\frac{1}{2}\sigma^{2}w^{2}+g_{r}\left(w\right)\right)+i\left(\frac{1}{2}\sigma^{2}w+g_{i}\left(w\right)\right)\right)\right),\label{eq:phi}
\end{align}
where $w\in\mathbb{R}$, $g\left(w\right)=f\left(w-i\right)-i\left(w-i\right)f\left(-i\right)$,
$g_{r}={\rm Re}\left(g\right)$, and $g_{i}={\rm Im}\left(g\right)$.
Note that $g\left(0\right)=g\left(i\right)=0$. By means of $\left[{\rm a}1^{**}\right]$
and $\left[{\rm a3}\right]$, we can show
\[
g\left(w\right)=h\left(w\right)-c_{0}-ic_{1}w,
\]
where 
\begin{equation}
h\left(w\right)=\int_{-\infty}^{\infty}e^{x}e^{iwx}\nu\left(dx\right),\;c_{0}=\int_{-\infty}^{\infty}e^{x}\nu\left(dx\right),\;c_{1}=\int_{-\infty}^{\infty}\left(e^{x}-1\right)\nu\left(dx\right).\label{eq:h}
\end{equation}
From the fact that $h$ is the Fourier transform of the real function
$e^{x}\frac{d\nu}{dx}\in L^{1}\left(\mathbb{R}\right)\cap L^{2}\left(\mathbb{R}\right)$,
the followings can be derived: $h\in L^{2}\left(\mathbb{R}\right)$,
$h\left(w\right)\rightarrow0$ as $\left|w\right|\rightarrow\infty$,
$h_{r}\left(-w\right)=h_{r}\left(w\right)$ and $h_{i}\left(-w\right)=-h_{i}\left(w\right)$
($h=h_{r}+ih_{i}$) (cf. \citet{pinsky2002introduction}). Additionally,
note that $h\left(0\right)=c_{0}$, $h\left(i\right)=c_{0}-c_{1}$,
and supposing that $h$ is differentiable, $h_{r}'\left(0\right)=0$.
Meanwhile, $\frac{d\nu}{dx}$ can be calculated from $h$ as follows:
\[
\frac{d\nu}{dx}\left(x\right)=e^{-x}\mathcal{F}^{-1}\left[h\left(w\right)\right]\left(x\right).
\]
It implies that finding $h$ enables us to estimate the L\'evy measure
$\nu$. Therefore, we try to derive $h$ from given data as well as
possible by utilizing ANNs elaborately. Strictly speaking, we use
feedforward ANNs. If unfamiliar with those, one can refer to preceding
papers, for examples, \citet{hutchinson1994nonparametric}.

We make two artificial neural networks (ANN) to approximate $h_{r}$
and $h_{i}$. More precisely,
\begin{align*}
h_{r}\left(w\right) & ={\rm ANN}_{r}\left(w;\theta_{r}\right),\;h_{i}\left(w\right)={\rm ANN}_{i}\left(w;\theta_{i}\right),
\end{align*}
where $\theta_{r}$ and $\theta_{i}$ mean the parameter sets of ${\rm ANN}_{r}$
and ${\rm ANN}_{i}$, respectively. The detailed structures of the
ANNs will be introduced after a while. By doing this, $\Phi_{X_{T}}\left(w-i\right)$
in (\ref{eq:phi}) is given by
\begin{align*}
\Phi_{X_{T}}\left(w-i;\Theta\right) & =\exp\left(T\left(\left(-\frac{1}{2}\sigma^{2}w^{2}+{\rm ANN}_{r}\left(w;\theta_{r}\right)-c_{0}\right)+i\left(\frac{1}{2}\sigma^{2}w+{\rm ANN}_{i}\left(w;\theta_{i}\right)-c_{1}w\right)\right)\right),
\end{align*}
where $\Theta=\left\{ c_{0},c_{1},\sigma,\theta_{r},\theta_{i}\right\} $.
Note that $\Phi_{X_{T}}\left(w-i\right)$ is a complex function. This
makes it difficult for the ANNs to learn market information because
only real-valued functions can be treated by existing learning methods
such as the stochastic gradient method. So, we decompose $\Phi_{X_{T}}\left(w-i\right)$
into two real-valued functions $\Phi_{X_{T},r}\left(w-i\right):={\rm Re}\left(\Phi_{X_{T}}\left(w-i\right)\right)$
and $\Phi_{X_{T},i}\left(w-i\right):={\rm Im}\left(\Phi_{X_{T}}\left(w-i\right)\right)$
with Euler's formula as follows:
\[
\Phi_{X_{T},r}\left(w-i\right)=\exp R\left(w\right)\cos\left({\rm Arg}\left(w\right)\right),\enskip\Phi_{X_{T},i}\left(w-i\right)=\exp R\left(w\right)\sin\left({\rm Arg}\left(w\right)\right),
\]
where
\[
R\left(w\right)=T\left(-\frac{1}{2}\sigma^{2}w^{2}+{\rm ANN}_{r}\left(w;\theta_{r}\right)-c_{0}\right),\enskip{\rm Arg}\left(w\right)=T\left(\frac{1}{2}\sigma^{2}w+{\rm ANN}_{i}\left(w;\theta_{i}\right)-c_{1}w\right).
\]

\begin{figure}
\centering{}\includegraphics[scale=0.5]{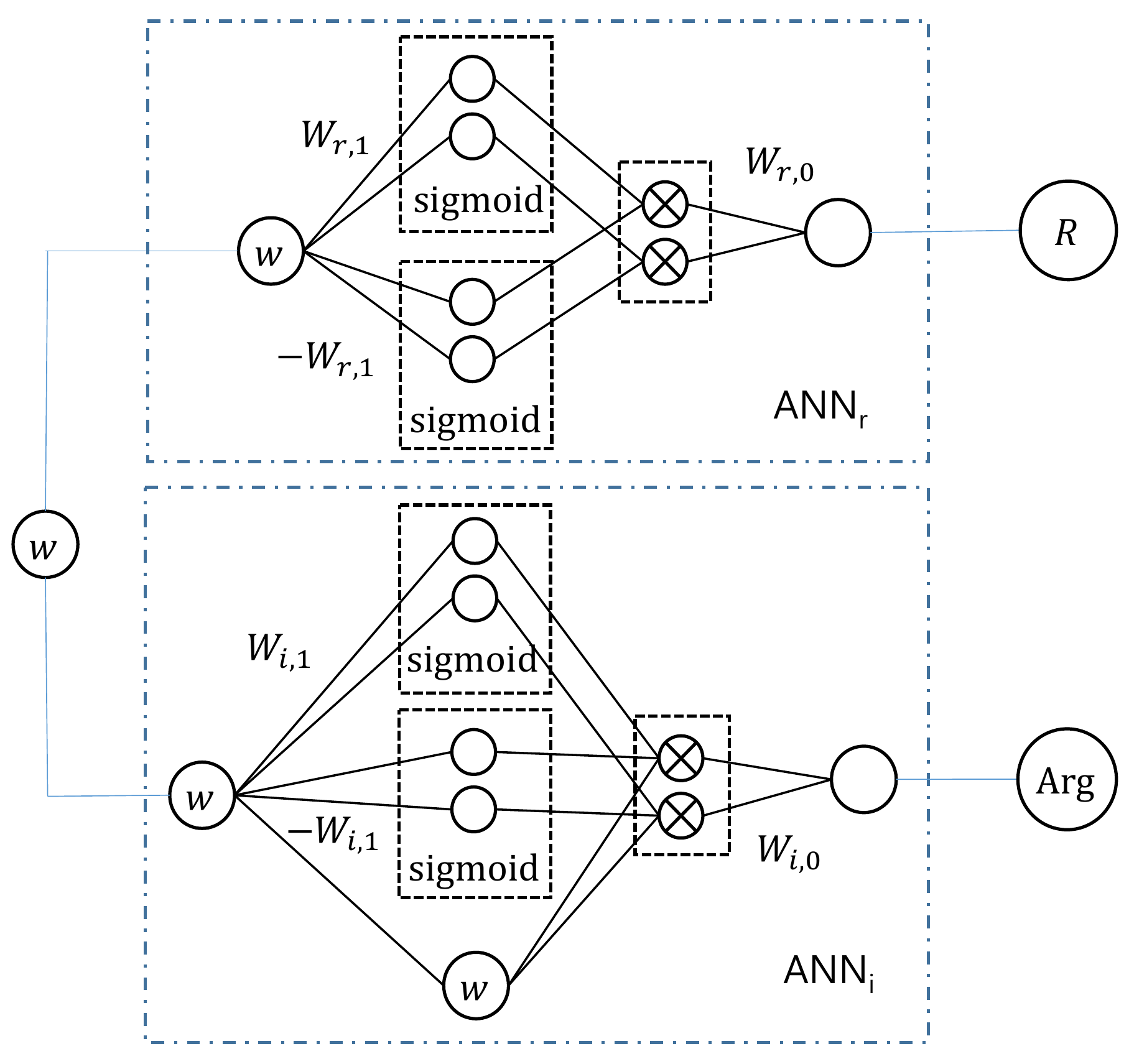}\caption{\label{fig:ELNN}This figure illustrates a part of the structure of
the exponential L\'evy neural network (ELNN). The nodes denoted by
$\varotimes$ produce the outputs given by multiplying all inputs
of each node.}
\end{figure}

In what follow, we design the ANNs so that they can inherit the properties
of $h$ as many as possible. So, it is desirable that the ANNs have
the following properties:
\begin{enumerate}
\item ${\rm ANN}_{r}\left(-w;\theta_{r}\right)={\rm ANN}_{r}\left(w;\theta_{r}\right)$
\item ${\rm ANN}_{r}\left(0;\theta_{r}\right)=c_{0}$
\item ${\rm ANN}_{r}'\left(0;\theta_{r}\right)=0$
\item ${\rm ANN}_{r}\left(w;\theta_{r}\right)\rightarrow0$ as $\left|w\right|\rightarrow\infty$
\item ${\rm ANN}_{i}\left(-w;\theta_{r}\right)=-{\rm ANN}_{i}\left(w;\theta_{r}\right)$
\item ${\rm ANN}_{i}\left(0;\theta_{r}\right)=0$
\item ${\rm ANN}_{r}\left(w;\theta_{r}\right)\rightarrow0$ as $\left|w\right|\rightarrow\infty$
\item ${\rm ANN}_{r}\left(i;\theta_{r}\right)+i{\rm ANN}_{i}\left(i;\theta_{i}\right)=c_{0}-c_{1}$
\end{enumerate}
After a careful consideration, we set
\[
{\rm ANN}_{r}\left(w\right)=W_{r,0}{\rm sig}\left(W_{r,1}w\right){\rm sig}\left(-W_{r,1}w\right),
\]
\[
{\rm ANN}_{i}\left(w\right)=W_{i,0}{\rm sig}\left(W_{i,1}w\right){\rm sig}\left(-W_{i,1}w\right)w,
\]
where $W_{r,0}$, $W_{r,1}$, $W_{i,0}$ and $W_{i,1}$ are weights,
and ${\rm sig}$ is the sigmoid function, i.e. ${\rm sig}\left(x\right)=1/\left(1+e^{-x}\right)$.
In addition, $c_{0}:={\rm ANN}_{r}\left(0;\theta_{r}\right)$ and
$c_{1}:={\rm ANN}_{r}\left(0;\theta_{r}\right)-{\rm ANN}_{r}\left(i;\theta_{r}\right)-i{\rm ANN}_{i}\left(i;\theta_{i}\right)$.
Using ${\rm sig}\left(\alpha i\right){\rm sig}\left(-\alpha i\right)=1/\left(2\left(1+\cos\alpha\right)\right)$,
$c_{0}$ and $c_{1}$ are explicitly expressed by
\[
c_{0}=\frac{1}{4}W_{r,0},\;c_{1}=\frac{1}{4}W_{r,0}-\frac{W_{r,0}}{2\left(1+\cos W_{r,1}\right)}+\frac{W_{i,0}}{2\left(1+\cos W_{i,1}\right)}.
\]
 Then, it is easy to show that the ANNs satisfy the foregoing properties
1,2,4,5,6,7,8. Because $\left({\rm sig}\left(w\right)\right)'={\rm sig}\left(w\right)\left(1-{\rm sig}\left(w\right)\right)$,
the derivatives of the ANNs are 
\[
{\rm ANN}_{r}'\left(w\right)=W_{r,0}{\rm sig}\left(W_{r,1}w\right){\rm sig}\left(-W_{r,1}w\right)\left(-W_{r,1}{\rm sig}\left(W_{r,1}w\right)+W_{r,1}{\rm sig}\left(-W_{r,1}w\right)\right),
\]
\[
{\rm ANN}_{i}'\left(w\right)=W_{i,0}{\rm sig}\left(W_{i,1}w\right){\rm sig}\left(-W_{i,1}w\right)\left(1+\left(-W_{i,1}{\rm sig}\left(W_{i,1}w\right)+W_{i,1}{\rm sig}\left(-W_{i,1}w\right)\right)w\right).
\]
Substituting $w=0$ into the above expressions, we can check that
the ANNs satisfy the property 3 also. 
\[
{\rm ANN}_{r}'\left(0\right)=0,\;{\rm ANN}_{i}'\left(0\right)=\frac{1}{4}W_{i,0}.
\]
The network whose input and output are $w$ and $\left(\Phi_{X_{T},r}\left(w-i\right),\Phi_{X_{T},i}\left(w-i\right)\right)$
respectively is hereafter called the exponential L\'evy neural network
(ELNN). By its great flexibility, if it is trained properly, the ELNN
can encompass and outperform the existing parametric exp-L\'evy models
with finite L\'evy measures such as the Merton and Kou models. Moreover,
in virtue of various studies for the optimization problems in neural
networks, it is plausibly better than the existing non-parametric
exp-L\'evy models such as the CT model and BR method. Figure \ref{fig:ELNN}
illustrates a part of the structure of the ELNN. The nodes denoted
by $\varotimes$ produce the outputs given by multiplying all inputs
of each node. At a glance, the ANNs of the ELNN may look similar to
neural networks with one hidden layer. But, in the ANNs proposed in
this paper, the nodes in the respective hidden layers are organized
into the two groups that contain the same number of elements. The
signals from the groups are merged into one signal as being matched
and multiplied. Moreover, the weights of the groups are closely related.
These elaborate designs make it possible to consider the ELNN as a
fully generalized version of the exp-L\'evy model.

Finally, we delineate a training method of the ELNN, i.e. a method
to find the optimal parameters $\Theta^{*}=\left\{ c_{0}^{*},c_{1}^{*},\sigma^{*},\theta_{r}^{*},\theta_{i}^{*}\right\} $.
As said earlier, our basic strategy is to locate $\Theta$ to minimize
$||\Phi_{X_{T}}\left(w-i;\Theta\right)-\Phi_{X_{T}}^{*}\left(w-i\right)||_{2}$.
However, naive applications of this approach are hard to carry out.
Recall that ${\rm ANN}_{r}\left(w\right)$ and ${\rm ANN}_{i}\left(w\right)$
are designed to go to 0 as $\left|w\right|\rightarrow\infty$. But,
in practice, it is still a difficult problem for the ANNs to guess
a proper constant $M>0$ such that their values are very small for
$\left|w\right|>M$. This problem outrageously increases the training
time for the ELNN. So, while observing the shape of $\Phi_{X_{T}}^{*}\left(w-i\right)$,
we manually set $M>0$ such that $|\Phi_{X_{T}}^{*}\left(w-i\right)|<N$
for $\left|w\right|>M$ and a certain value $N>0$. With this $M$,
a regularization function $\Lambda$ is introduced:
\begin{equation}
\Lambda\left(\theta_{r},\theta_{i}\right)=\int_{-\infty}^{\infty}\left|\frac{w}{M}\right|^{\alpha}\left({\rm ANN}_{r}\left(w;\theta_{r}\right){}^{2}+{\rm ANN}_{i}\left(w;\theta_{i}\right)^{2}\right)dw,\label{eq:Lambda}
\end{equation}
where $\alpha$, at least larger than $1$, is set to be $4$ in this
paper. We find that penalizing the original objective function $||\Phi_{X_{T}}\left(w-i;\Theta\right)-\Phi_{X_{T}}^{*}\left(w-i\right)||_{2}$
with $\Lambda$ fairly reduces the training time. Note that $\left|w/M\right|^{\alpha}$
in $\Lambda$ is very small for $\left|w\right|\ll M$, while it is
very large for $\left|w\right|\gg M$. This allows us to compress
the ANNs into $0$ for $\left|w\right|\gg M$ without affecting them
for $\left|w\right|\ll M$. In summary, the objective function for
the ELNN is
\[
||\Phi_{X_{T}}\left(w-i;c_{0},c_{1},\sigma,\theta_{r},\theta_{i}\right)-\Phi_{X_{T}}^{*}\left(w-i\right)||_{2}^{2}+\beta\Lambda\left(\theta_{r},\theta_{i}\right),
\]
where $\beta$ is a small constant, and we set this value as $0.001$.
Because each term of this function is associated with an integration
on $\mathbb{R}$, we discretize it on an enough large compact set
by using the trapezoid rule.

\section{\label{sec:Numerical-tests}Numerical tests}

\begin{figure}[t]
\begin{centering}
\subfloat[the Merton model]{\centering{}\includegraphics[scale=0.62]{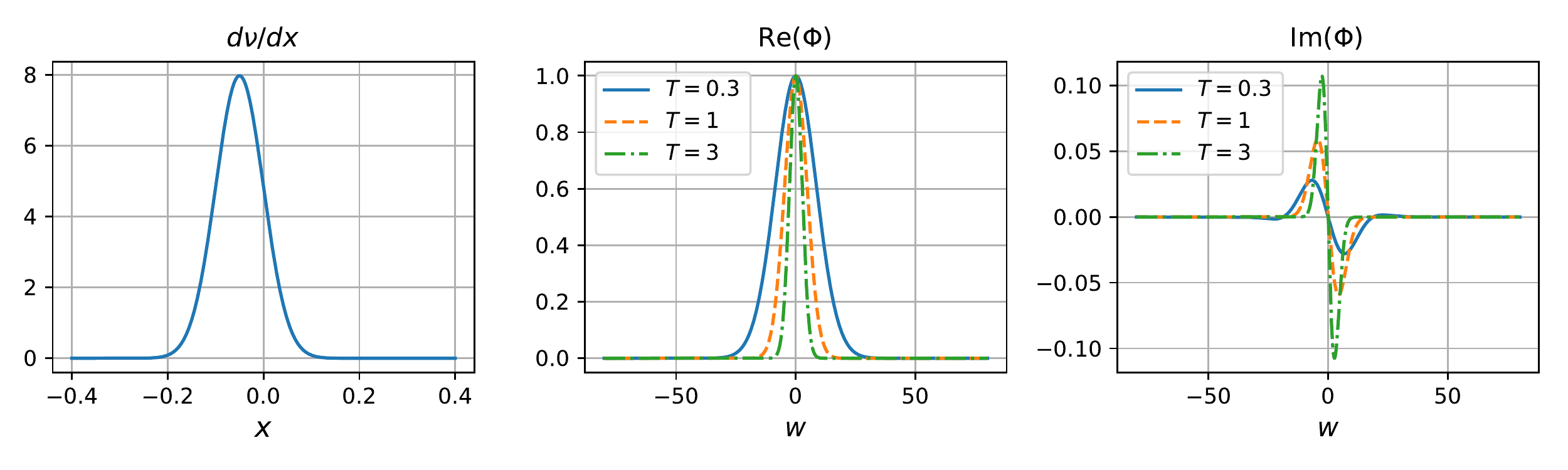}}
\par\end{centering}
\centering{}\subfloat[the Kou model]{\centering{}\includegraphics[scale=0.62]{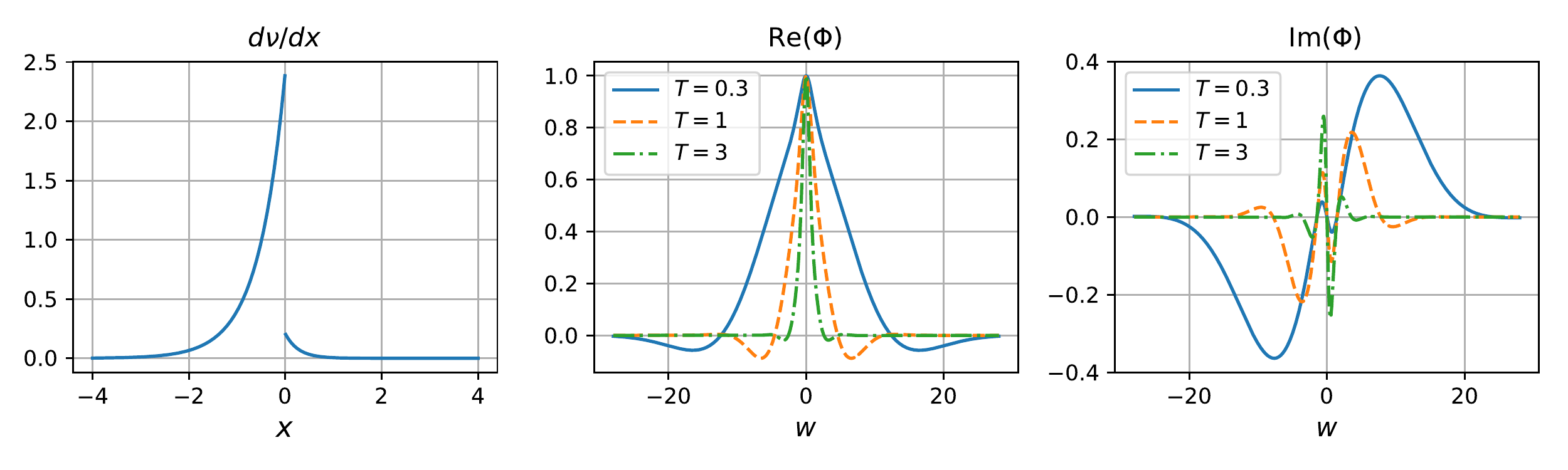}}\caption{\label{fig:characteristics} This figure depicts $\frac{d\nu}{dx}$
and $\Phi_{X_{T}}\left(w-i\right)$ on three maturities $T$ for the
Merton and Kou models. The parameter values for the models are written
in the text.}
\end{figure}

In this section, we implement the ELNN in the previous section. To
this end, we generate 100,000 virtual option prices $\tilde{c}_{X}^{*}\left(k\right)$
on various $k$ over 1,000 days (100 options per day) by employing
two parametric exp-L\'evy models, the Merton and Kou models. Recall
that these models are based on finite L\'evy measures. Exp-L\'evy
models with infinite L\'evy measures such as the CGMY model is not
in interest of this paper because the ELNN is not designed for the
cases. Figure \ref{fig:characteristics} describes the L\'evy densities
$\frac{d\nu}{dx}$ and $\Phi_{X_{T}}\left(w-i\right)$ on $r=0.02$
and $T=\left\{ 0.3,1,3\right\} $ for the Merton and Kou models. Their
parameters are set as follows: $\sigma=0.2$, $\lambda=1$, $\mu=-0.05$
and $\delta=0.05$ in the Merton model (\ref{eq:merton_density}),
and $\sigma=0.21$, $p=0.04$, $\lambda=1.4$, $\lambda_{+}=3.7$
and $\lambda_{-}=1.8$ in the Kou model (\ref{eq:kou_density}). We
choose proper parameter sets by monitoring sample paths from a number
of Monte-Carlo simulations. In the figure, the L\'evy densities of
the two models have distinct shapes. The graph for the Merton model
is continuous, whereas the one for the Kou model is discontinuous.
Recall that \citet{cont2004nonparametric} directly estimated L\'evy
densities. However, this method can cause severe instability when
estimating discontinuous densities such as the case of the Kou model.
Regarding the ELNN, the problem does not occur because the ANNs in
the ELNN approximate the continuous $h$ in (\ref{eq:h}). It is also
interesting to observe the shapes of $\Phi_{X_{T}}\left(w-i\right)$
according to $T$: the shorter the maturity is, the more clearly $\Phi_{X_{T}}\left(w-i\right)$
depends on model types. But they become similar to each other at long
maturities, which closely resemble the characteristic functions of
normal distributions. This is because a lot of characteristics from
jumps fade due to the central limit theorem. So, when estimating the
L\'evy measure for an asset, it is effective to exploit short-term
options on the asset. Moreover, as aforementioned in Section \ref{sec:exp_levy_models},
the exp-L\'evy models are suitable at a fixed short time horizon.
For these reasons, we process this experiment only for the short maturity
$T=0.05$.

\begin{sidewaysfigure}
\begin{centering}
\subfloat[\label{fig:Merton}the Merton model]{\begin{centering}
\includegraphics[scale=0.62]{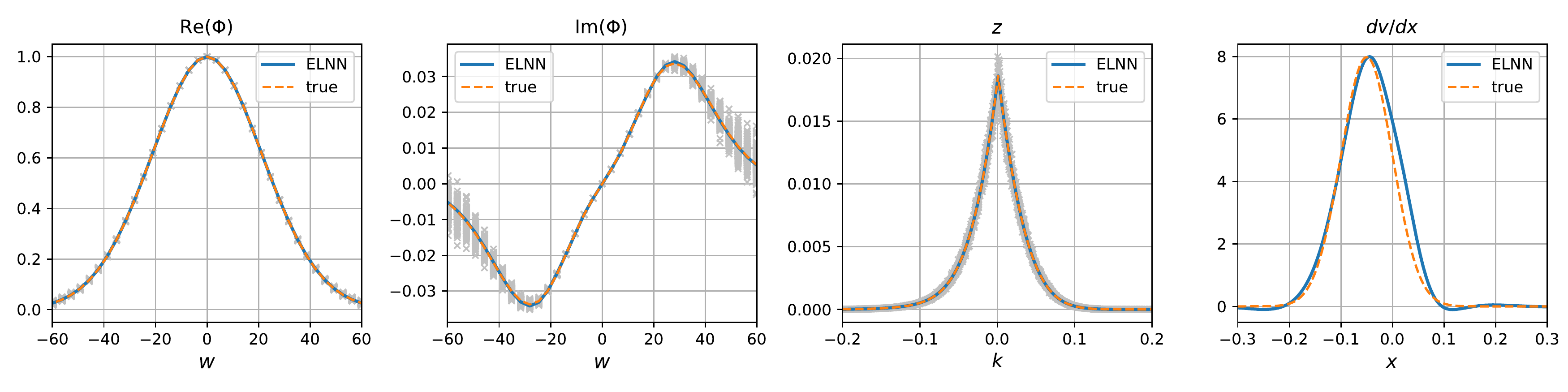}
\par\end{centering}
}
\par\end{centering}
\subfloat[\label{fig:kou}the Kou model]{\begin{centering}
\includegraphics[scale=0.62]{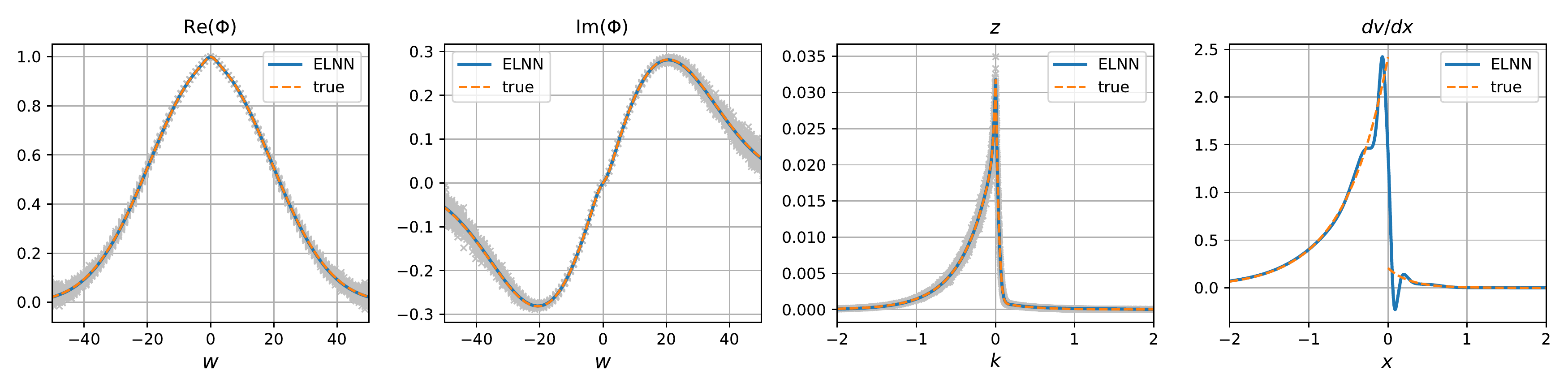}
\par\end{centering}
}\caption{\label{fig:the-fitting-results}This figure graphs $\Phi_{X_{T}}\left(w-i\right)$,
$z_{X_{T}}$ and $\frac{dv}{dx}$ for the Merton and Kou models. The
gray markers mean $z_{X_{T}}^{*}$ or $\Phi_{X_{T}}^{*}\left(w-i\right)$
computed from $z_{X_{T}}^{*}$.}
\end{sidewaysfigure}

We then calculate $z_{X_{T}}^{*}$ from $\tilde{c}_{X}^{*}$ and add
market noises with $N(0,0.05z_{X_{T}}^{*})$ to $z_{X_{T}}^{*}$ as
in the work of \citet{belomestny2006spectral}. For convenience, we
denote the perturbed $z_{X_{T}}^{*}$ by $z_{X_{T}}^{*}$ without
changing the notation. It seems reasonable that a market noise on
an option price becomes stronger as the time value of the option gets
larger. Next, we obtain $\Phi_{X_{T}}^{*}\left(w-i\right)$ using
(\ref{eq:phi*}) for each day and let the ELNN learn them, thereby
obtaining $\Phi_{X_{T}}\left(w-i\right)$ predicted by the ELNN. In
the learning process, the number of nodes per group is set to be 20,
so the total number of nodes in the ELNN is 80. We find that more
nodes do not necessarily guarantee better results. The regularization
parameter $M$ in (\ref{eq:Lambda}) is set as follows: $M=100$ for
the virtual market by the Merton model, and $M=55$ for the Kou market.
In addition, the ADAM optimizer (\citet{kingma2014adam}) is used
to find a good local minimum of the objective function proposed in
the previous section. Various optimizers are tested, but the ADAM
optimizer is superior to the others, at least for the ELNN. The learning
is processed during 30,000 epochs with full batches. In fact, we use
a trick ``data amplification'' when obtaining $\Phi_{X_{T}}^{*}\left(w-i\right)$
from $z_{X_{T}}^{*}$ for each day because 100 points in $k$-space
are not enough to find quite an accurate transform in $w$-space.
With the data gathering all $z_{X_{T}}^{*}$ over the whole data period
(1,000 days), we make 1,000 groups such that each of them includes
10,000 randomly chosen $z_{X_{T}}^{*}$. We then regard each group
as individual one-day data. Compared to the initial setting, it can
be viewed as amplifying the original data 100 times for each day.
Notice that this approach is plausible because $z_{X_{T}}^{*}$ does
not rely on $S_{0}$. It will prove very effective after a little.

Figure \ref{fig:the-fitting-results} graphs $\Phi_{X_{T}}\left(w-i\right)$,
$z_{X_{T}}$ and $\frac{dv}{dx}$ for the Merton and Kou models. The
blue solid lines and the dashed orange lines indicate the predicted
values by the ELNN and the true values for the models, respectively.
The gray markers mean $z_{X_{T}}^{*}$ or $\Phi_{X_{T}}^{*}\left(w-i\right)$
computed from $z_{X_{T}}^{*}$. Because the market noises perturb
the option prices, $\Phi_{X_{T}}^{*}\left(w-i\right)$ often deviates
from correct values particularly when $\left|w\right|$ is large,
which greatly increases the likelihood of a biased learning. But,
by training the ELNN on the long-term data, we make the errors cancel
each other out and reduce the risk of bias. Note that the prediction
lines successfully pass through the given data points. Additionally,
these prediction values are very close to exact values, which verifies
that the ELNN has an outstanding reasoning ability robust to market
noises. On the other hand, the ELNN gives a little worse results when
it comes to $\frac{dv}{dx}$ near $x=0$ for the Kou model. This is
caused by the Gibbs phenomenon, which arises from that the L\'evy
density of the Kou model is discontinuous at $x=0$. The ELNN estimates
the volatilities $\sigma$ as 0.1998 for the Merton model and 0.2106
for the Kou model. These values have only 0.1153\% and 0.2791\% relative
errors, respectively.

\begin{figure}[p]
\begin{centering}
\subfloat[$\sigma=0.2$]{\centering{}\includegraphics[scale=0.62]{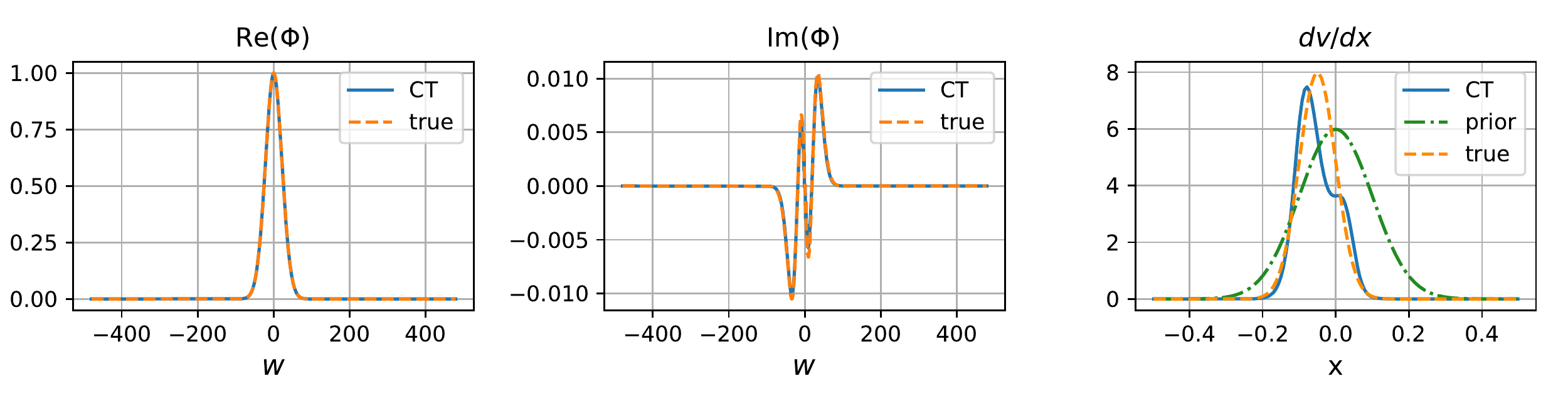}}
\par\end{centering}
\centering{}\subfloat[$\sigma=0.195$]{\centering{}\includegraphics[scale=0.62]{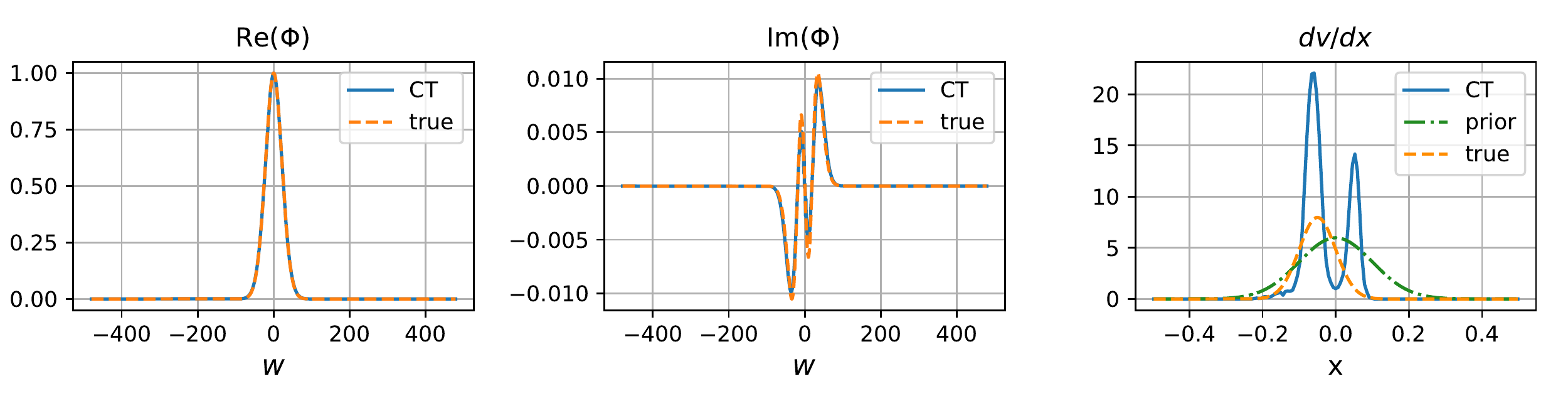}}\caption{\label{fig:CT}This figure represents $\Phi_{X_{T}}\left(w-i\right)$
and $\frac{dv}{dx}$ for the CT model. The upper and lower figures
are the cases where $\sigma$ of priors are set to be 0.2 and 0.195,
respectively. For one's information, $\sigma=0.2$ in the virtual
market.}
\end{figure}

\begin{figure}[p]
\begin{centering}
\subfloat[\label{fig:BR-no-fft}the cases not allowing FFT errors]{\centering{}\includegraphics[scale=0.62]{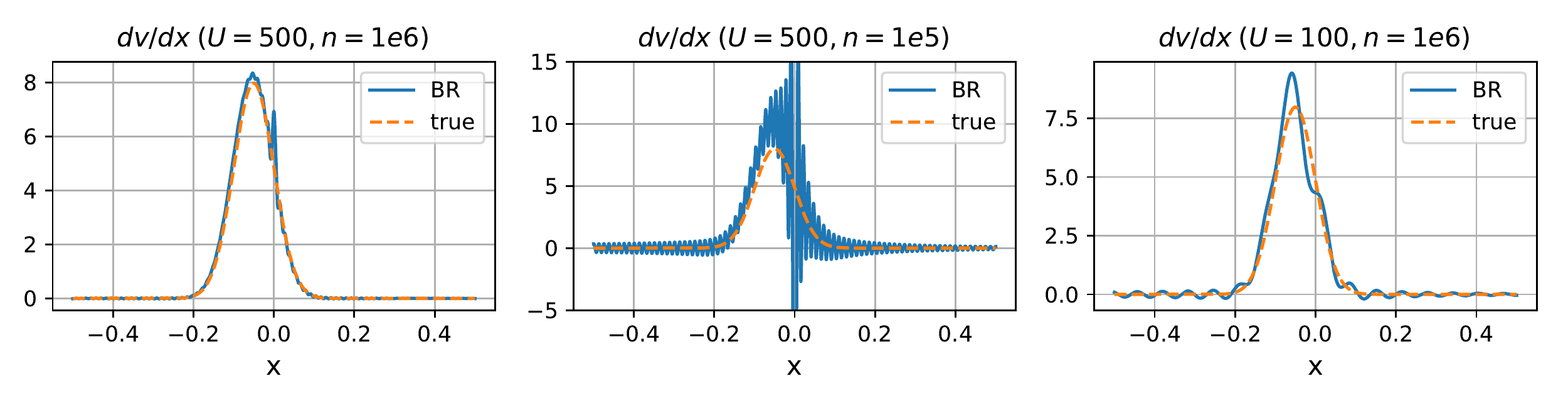}}
\par\end{centering}
\centering{}\subfloat[\label{fig:BR-fft}the cases allowing FFT errors]{\begin{centering}
\includegraphics[scale=0.62]{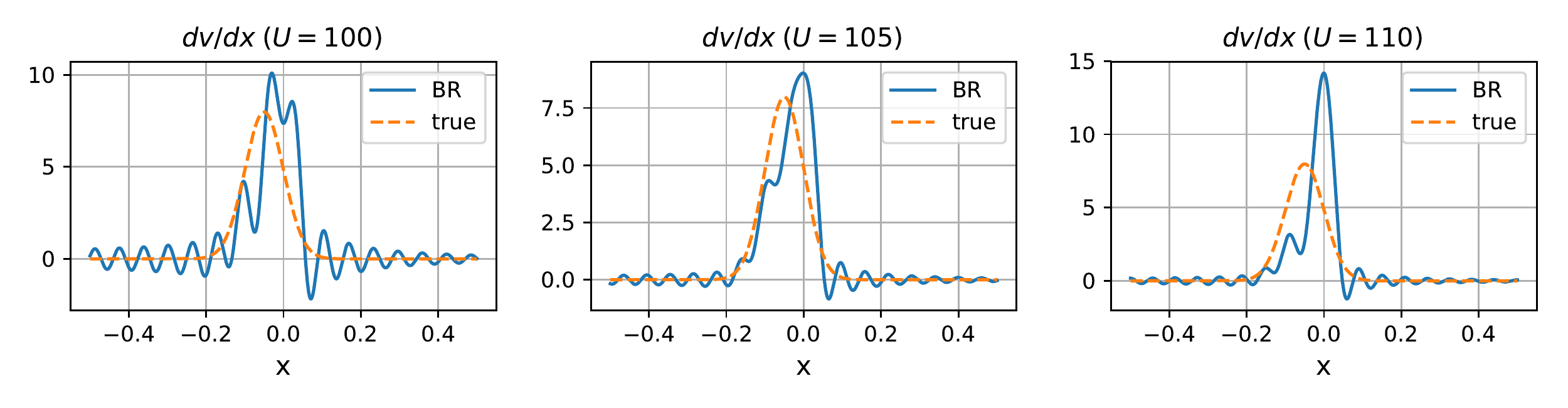}
\par\end{centering}
}\caption{This figure shows the instability of the BR method by depicting $\frac{dv}{dx}$
inferred under various conditions. Here, $U$ is a cutoff value, and
$n$ is the number to partition $\left[-U,U\right]$.}
\end{figure}

In order to clarify our contribution to literature, we should compare
the ELNN with the other non-parametric exp-L\'evy models, the CT
model and the BR method. However, it is already known that the other
models are so vulnerable to various factors that they are practically
demanding to adopt. So, we analyze and reveal the vulnerabilities
instead of the direct comparison with the ELNN. This additional test
is made on the existing data by the Merton model, among which we use
its only one-day data not including the market noises. At first, let
us look into the CT model. As mentioned previously, Cont and Tankov
put a penalty term as the relative entropy with respect to a Bayesian
prior. A Bayesian prior means the L\'evy measure of a L\'evy process,
which acts as an initial guess for optimization. But this regularization
makes calibration results depend on the prior considerably. For this,
the CT model can not help getting the same $\sigma$ as that of the
prior. Thus, setting $\sigma$ of the prior different from the actual
value gives rise to an inexact estimation of the L\'evy density.
One can check these facts through Figure \ref{fig:CT} to represent
$\Phi_{X_{T}}\left(w-i\right)$ and $\frac{dv}{dx}$ for the CT model.
The upper and lower figures correspond to the cases where $\sigma$
of priors are set to be 0.2 and 0.195, respectively. Recall that $\sigma$
is set to be 0.2 in the Merton model. Apart from this, it is interesting
that the shapes of $\frac{dv}{dx}$ are very distinct although both
the cases give quite exact $\Phi_{X_{T}}\left(w-i\right)$. It implies
that estimating L\'evy measures from option prices is ill-conditioned,
so it should be handled carefully. What follows is to examine the
BR method. Once the assumptions for the method are satisfied completely,
i.e. the spectral cutoff $U$ and the number $n$ to partition $\left[-U,U\right]$
are large enough, it works as expected. But, if any one of the assumptions
fails, its performance gets greatly compromised. Figure \ref{fig:BR-no-fft}
proves them. In fact, the figure describes the cases where FFT errors
do not exist by using known true values in $w$-space without the
FFT. If the errors are allowed, the BR method gives significantly
different results according to $U$ as in Figure \ref{fig:BR-fft}.

\section{\label{sec:Empirical-tests}Empirical tests}

To evaluate the outperformance of the ELNN under actual markets, we
conduct empirical tests to compare it with two existing exp-L\'evy
models: the Merton and Kou models. Here, we exclude various models
using ANNs and the existing non-parametric exp-L\'evy models from
this analysis. Note that the ELNN is designed to belong to a category
of strong hybrid models so as to avoid the essential issues of the
existing ANN-based models. Supposing that the non-hybrid models or
the weak hybrid models are superior to the ELNN in terms of goodness
of fit, the models with the drawbacks can not be alternatives of the
ELNN (Refer to the introduction). To the best of our knowledge, \citet{yang2017gated}
have, until now, suggested the only strong hybrid model for option
pricing except for the ELNN. But it is difficult to apply general
analysis of our study to the model because it can not be considered
in view of the frequency domain. So, we leave a comparative study
between the model and the ELNN as a further research so as to keep
consistency of this paper's context. On the other hand, the existing
non-parametric exp-L\'evy models, the CT model and the BR method,
are too vulnerable to be employed in practice, as indicated in the
preceding section. 

\begin{figure}
\begin{centering}
\includegraphics[scale=0.62]{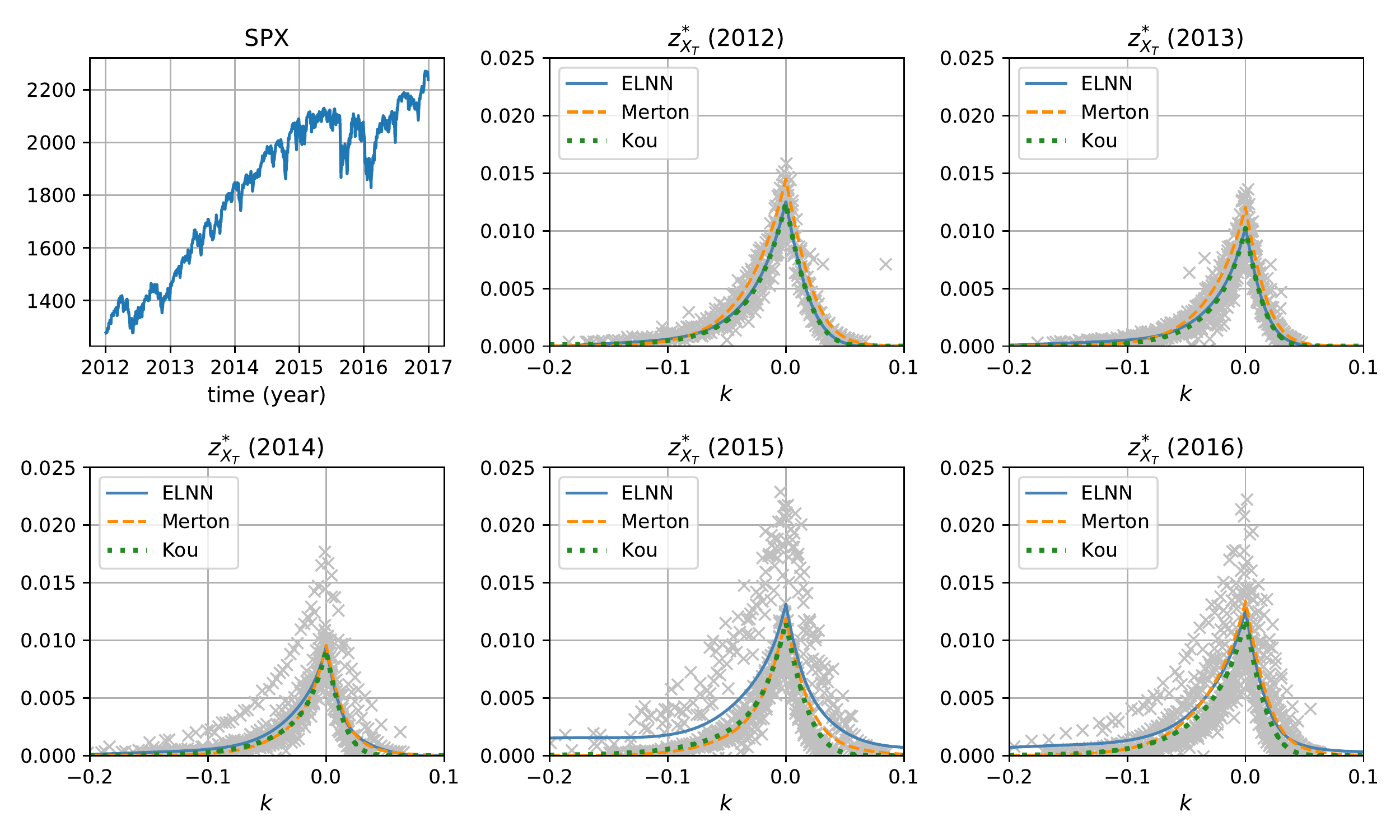}
\par\end{centering}
\caption{\label{fig:emp-z}This figure describes $z_{X_{T}}^{*}$ from SPX
options and the lines predicted by fitting three exp-L\'evy models
to them. The gray markers indicate $z_{X_{T}}^{*}$. The data period
ranges from 2012 to 2016, and the SPX series during the period are
drawn in the upper left subfigure.}
\end{figure}

For this test, we have used the market data of the Chicago Board Options
Exchange (CBOE) for 5 years from 2012 to 2016.\footnote{CBOE DataShop (https://datashop.cboe.com)}
After filtering out some options if they are traded lower than 100
times in a day or their prices are below 0.5, we choose a short time-to-maturity
$T=13/252$ and collect the corresponding closing prices of 1,273
calls and 2,054 puts on the S\&P 500 index (SPX). Each put price of
the data is converted to a call price by the put-call parity, and
the whole data is then divided into 5 non-overlapping one-year subperiods.
From these prices $\tilde{c}_{X}^{*}$, $z_{X_{T}}^{*}$ and $\Phi_{X_{T}}^{*}\left(w-i\right)$
have been computed for each year. The data amplification in Section
\ref{sec:Numerical-tests} are utilized also here. Figure \ref{fig:emp-z}
draws $z_{X_{T}}^{*}$ among them with gray markers. The fitting results
in the figure will be explained in a while. One can note that the
prices are spreaded more widely for 2015 and 2016 compared to the
other periods. This is because the data period includes the Chinese
stock market crash (from June 2015 to February 2016). Observe that
the SPX series dramatically change around the crisis as shown in the
upper left of the figure. In fact, the wide distribution are made
because the values of $z_{X_{T}}^{*}$ move to larger values during
the crisis and revert to the original ones after the time. This is
an obvious evidence of heteroscedasticity, which can not be explained
with the exp-L\'evy models. But, as said before, we set aside the
problem and focus on a fully generalization of the exp-L\'evy models.
Thus, we just think that the wide-spread prices are generated by large
market noises. In addition, we average one-month treasury bill yields
for each year and use them as risk-free rates.

\begin{table}[p]
\begin{centering}
\subfloat[errors for $z_{X_{T}}^{*}$]{%
\begin{tabular}{ccccccccccccc}
\hline 
 & \multicolumn{4}{c}{2012} & \multicolumn{4}{c}{2013} & \multicolumn{4}{c}{2014}\tabularnewline
 & ATM & ITM & OTM & sum & ATM & ITM & OTM & sum & ATM & ITM & OTM & sum\tabularnewline
\hline 
ELNN & \textbf{14.8} & \textbf{6.3} & \textbf{11.4} & \textbf{32.5} & \textbf{15.4} & \textbf{6.1} & \textbf{10.5} & \textbf{32.0} & \textbf{21.5} & \textbf{8.8} & \textbf{11.8}  & \textbf{42.1}\tabularnewline
Merton & 20.4 & 6.4 & 12.8 & 39.5 & 19.4 & 6.7 & 10.7 & 36.8 & 21.9 & 9.9 & 12.0 & 43.8\tabularnewline
Kou & 15.0 & 6.8 & 11.5 & 33.3 & 15.5 & 6.7 & 10.8 & 33.0 & 22.3 & 9.4 & 14.1 & 45.8\tabularnewline
\hline 
 & \multicolumn{4}{c}{2015} & \multicolumn{4}{c}{2016} & \multicolumn{4}{c}{average}\tabularnewline
 & ATM & ITM & OTM & sum & ATM & ITM & OTM & sum & ATM & ITM & OTM & sum\tabularnewline
\hline 
ELNN & \textbf{42.4} & 21.4 & 21.6 & \textbf{85.4} & \textbf{36.5} & 16.9 & \textbf{13.7} & \textbf{67.2} & \textbf{26.1} & \textbf{11.9} & \textbf{13.8} & \textbf{51.9}\tabularnewline
Merton & 43.4 & 22.2 & \textbf{21.1} & 86.7 & 37.2 & 16.5 & 13.8 & 67.5 & 28.5 & 12.3 & 14.1 & 54.9\tabularnewline
Kou & 43.6 & \textbf{21.2} & 24.1 & 88.9 & 36.7 & \textbf{16.4} & 15.8 & 68.9 & 26.6 & 12.1 & 15.3 & 54.0\tabularnewline
\hline 
 &  &  &  &  &  &  &  &  &  &  &  & \tabularnewline
\end{tabular}

}
\par\end{centering}
\begin{centering}
\subfloat[errors for ${\rm {\rm Re}}(\Phi_{X_{T}}^{*})$]{\begin{centering}
\begin{tabular}{ccccccccccccc}
\hline 
 & \multicolumn{4}{c}{2012} & \multicolumn{4}{c}{2013} & \multicolumn{4}{c}{2014}\tabularnewline
 & Low & Mid & High & sum & Low & Mid & High & sum & Low & Mid & High & sum\tabularnewline
\hline 
ELNN & 6.4 & \textbf{12.9} & 41.3 & \textbf{60.5} & 9.1 & \textbf{15.0} & 35.3 & 59.3 & 10.3 & \textbf{22.7} & 75.3 & \textbf{108.3}\tabularnewline
Merton & \textbf{5.5} & 17.1 & \textbf{39.6} & 62.2 & 6.3 & 21.7 & \textbf{33.5} & 61.5 & \textbf{9.9} & 25.0 & \textbf{74.4} & 109.3\tabularnewline
Kou & 5.9 & 12.9 & 42.6 & 61.4 & \textbf{5.3} & 15.2 & 35.9 & \textbf{56.4} & 22.4 & 37.6 & 86.9 & 146.9\tabularnewline
\hline 
 & \multicolumn{4}{c}{2015} & \multicolumn{4}{c}{2016} & \multicolumn{4}{c}{average}\tabularnewline
 & Low & Mid & High & sum & Low & Mid & High & sum & Low & Mid & High & sum\tabularnewline
\hline 
ELNN & \textbf{18.3} & \textbf{31.4} & \textbf{87.8} & \textbf{137.5} & 12.2 & \textbf{28.5} & \textbf{69.4} & \textbf{110.2} & \textbf{11.3} & \textbf{22.1} & 61.8 & \textbf{95.2}\tabularnewline
Merton & 30.1 & 34.1 & 87.9 & 152.1 & 16.7 & 28.9 & 70.0 & 115.6 & 13.7 & 25.4 & \textbf{61.1} & 100.1\tabularnewline
Kou & 20.8 & 64.9 & 104.9 & 190.6 & \textbf{10.7} & 34.0 & 70.4 & 115.0 & 13.0 & 32.9 & 68.1 & 114.1\tabularnewline
\hline 
 &  &  &  &  &  &  &  &  &  &  &  & \tabularnewline
\end{tabular}
\par\end{centering}
\centering{}}
\par\end{centering}
\begin{centering}
\subfloat[errors for ${\rm {\rm Im}}(\Phi_{X_{T}}^{*})$]{\begin{centering}
\begin{tabular}{ccccccccccccc}
\hline 
 & \multicolumn{4}{c}{2012} & \multicolumn{4}{c}{2013} & \multicolumn{4}{c}{2014}\tabularnewline
 & Low & Mid & High & sum & Low & Mid & High & sum & Low & Mid & High & sum\tabularnewline
\hline 
ELNN & \textbf{7.3} & \textbf{10.2} & \textbf{14.3} & \textbf{31.9} & \textbf{7.9} & \textbf{9.4} & \textbf{11.5} & \textbf{28.7} & \textbf{10.8} & \textbf{18.2} & \textbf{22.0} & \textbf{51.1}\tabularnewline
Merton & 66.0 & 22.9 & 15.2 & 104.1 & 63.3 & 21.3 & 12.6 & 97.3 & 67.4 & 24.6 & 22.9 & 114.9\tabularnewline
Kou & 56.0 & 18.1 & 16.6 & 90.7 & 55.4 & 16.9 & 13.1 & 85.4 & 55.2 & 21.0 & 22.6 & 98.8\tabularnewline
\hline 
 & \multicolumn{4}{c}{2015} & \multicolumn{4}{c}{2016} & \multicolumn{4}{c}{average}\tabularnewline
 & Low & Mid & High & sum & Low & Mid & High & sum & Low & Mid & High & sum\tabularnewline
\hline 
ELNN & \textbf{22.0} & \textbf{29.0} & \textbf{43.1} & \textbf{94.1} & \textbf{11.6} & \textbf{16.8} & 25.3 & \textbf{53.7} & \textbf{11.9} & \textbf{16.7} & \textbf{23.2} & \textbf{51.9}\tabularnewline
Merton & 74.1 & 32.9 & 44.4 & 151.4 & 55.9 & 19.7 & 25.4 & 100.9 & 65.4 & 24.3 & 24.1 & 113.7\tabularnewline
Kou & 59.7 & 29.6 & 43.3 & 132.7 & 46.4 & 17.7 & \textbf{25.2} & 89.3 & 54.5 & 20.7 & 24.2 & 99.4\tabularnewline
\hline 
 &  &  &  &  &  &  &  &  &  &  &  & \tabularnewline
\end{tabular}
\par\end{centering}
\centering{}}
\par\end{centering}
\bigskip{}

\centering{}%
\begin{tabular}{ccc}
\hline 
ATM & ITM & OTM\tabularnewline
\hline 
$-0.05\leq k<0.03$ & $k<-0.05$ & $k\geq0.03$\tabularnewline
\hline 
 &  & \tabularnewline
\end{tabular}\qquad{}%
\begin{tabular}{ccc}
\hline 
Low & Mid & High\tabularnewline
\hline 
$w<20$ & $20\leq w<40$ & $40\leq w<60$\tabularnewline
\hline 
 &  & \tabularnewline
\end{tabular}\caption{\label{tab:in-sample-errors}This table summarizes the results of
in-sample tests on three exp-L\'evy models, which are separately
performed for each year. The smallest error of each group is highlighted
in bold.}
\end{table}

Next, we separately perform in-sample tests for each year with the
three exp-L\'evy models, the ELNN, the Merton and Kou models. When
training the ELNN, most conditions in the previous section are kept.
But the regularization parameter $M$ and the number of epochs are
changed to 65 and 100,000, respectively. For consistent fitting, the
parameters of the Merton and Kou models are also estimated by minimizing
$||\Phi_{X_{T}}\left(w-i\right)-\Phi_{X_{T}}^{*}\left(w-i\right)||_{2}$,
where $\Phi_{X_{T}}\left(w-i\right)$ are model-predicted values.
Because the closed-form expressions for $\Phi_{X_{T}}\left(w-i\right)$
are known under the parametric models, the optimizations are relatively
simple (cf. \citet{tankov2003financial}). 

The test results are summarized in Table \ref{tab:in-sample-errors},
and Figure \ref{fig:emp-z} visualizes them. The figures in the table
are displayed by groups to observe the effect of the ELNN more closely.
The errors for $z_{X_{T}}^{*}(=z_{X_{T}}^{*}\left(k\right))$ are
classified by $k$ into three groups: at-the-money (ATM), in-the-money
(ITM) and out-of-the-money (OTM) groups. The errors for $\Phi_{X_{T}}^{*}\left(w-i\right)$
are grouped by $w$ into three classes: low-frequency (Low), medium-frequency
(Mid) and high-frequency (High) groups. The errors are the root-mean-square
deviations (RMSE). To improve readability, the errors for $z_{X_{T}}^{*}$
are multiplied by 10,000, and the errors for $\Phi_{X_{T}}^{*}\left(w-i\right)$
are multiplied by 100 and divided by the standard deviation of each
group. The deviations are used to adjust the scales of the groups
associated with the errors for $\Phi_{X_{T}}^{*}\left(w-i\right)$
because $\Phi_{X_{T}}^{*}\left(w-i\right)$ becomes dispersed too
widely as $\left|w\right|$ increases. In addition, the smallest error
of each group is highlighted in bold. Now, we examine the values for
$z_{X_{T}}^{*}$. Judging from the averaged sums, hereafter called
the overall errors, the ELNN reduces 5.46\% and 3.89\% of the overall
errors for the Merton and Kou models, respectively. This is because
the parametric models often fail to give proper values to all the
groups. The models are not flexible enough to fit the data points
of all the groups nicely because their decay rates of $z_{X_{T}}$
are theoretically fixed. On the contrary, the ELNN can be adapted
for various distributions of $z_{X_{T}}^{*}$. The improvement becomes
more clear when it comes to the errors for $\Phi_{X_{T}}^{*}\left(w-i\right)$,
particularly for its imaginary part. As for ${\rm Re}(\Phi_{X_{T}}^{*}\left(w-i\right))$,
the ELNN decreases 4.90\% and 16.56\% of the overall errors for the
Merton and Kou models, respectively. Regarding ${\rm Im}(\Phi_{X_{T}}^{*}\left(w-i\right))$,
the overall error of the ELNN is as much as 54.35\% and 47.79\% smaller
than those of the Merton and Kou models, respectively. This is because
the parametric models are not good at describing ${\rm Im}(\Phi_{X_{T}}^{*}\left(w-i\right))$
of the low frequency group. In the table, note that the relevant values
are relatively large under the models.

\begin{figure}
\begin{centering}
\includegraphics[scale=0.62]{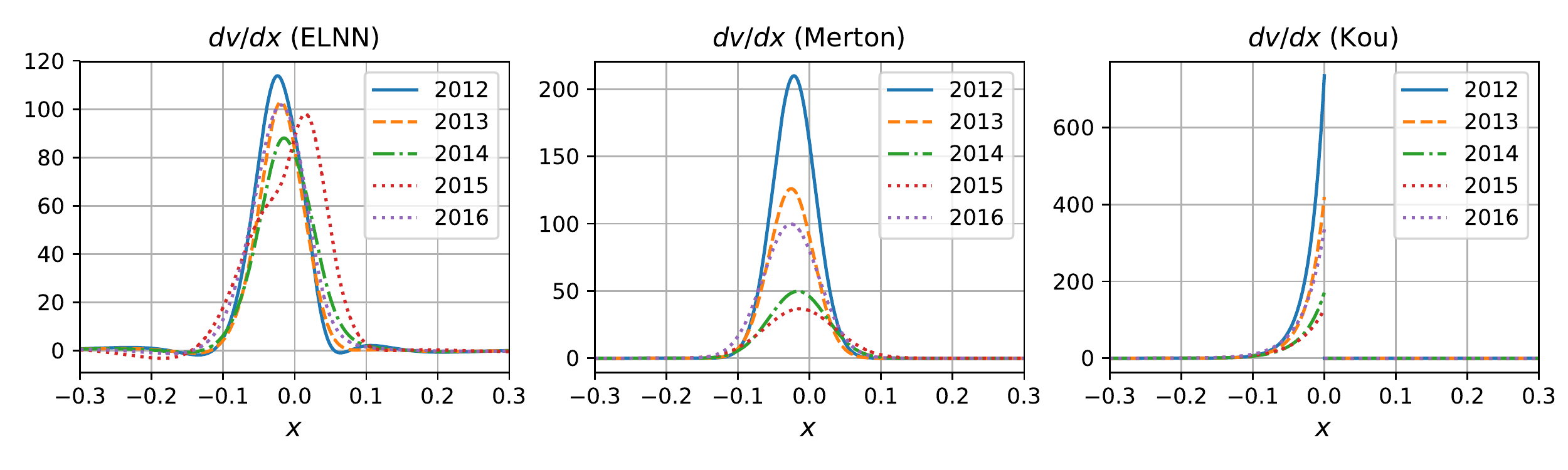}
\par\end{centering}
\caption{\label{fig:implied_density}This figure draws the L\'evy densities
$dv/dx$ estimated by deploying the models.}
\end{figure}

\begin{table}
\centering{}\subfloat[the estimates of $\sigma$]{\begin{centering}
\begin{tabular}{cccccc}
\hline 
 & 2012 & 2013 & 2014 & 2015 & 2016\tabularnewline
\hline 
ELNN & 8.95 & 6.88 & 4.79 & 5.96 & 6.80\tabularnewline
Merton & 8.17 & 7.11 & 5.56 & 6.81 & 7.05\tabularnewline
Kou & 8.83 & 7.32 & 7.32 & 9.89 & 7.81\tabularnewline
\hline 
\end{tabular}
\par\end{centering}
}\enskip{}\subfloat[the estimates of $\lambda$]{\centering{}%
\begin{tabular}{cccccc}
\hline 
 & 2012 & 2013 & 2014 & 2015 & 2016\tabularnewline
\hline 
ELNN & 9.06 & 7.91 & 8.27 & 10.62 & 9.49\tabularnewline
Merton & 11.19 & 6.66 & 6.95 & 9.03 & 8.30\tabularnewline
Kou & 15.59 & 9.82 & 5.01 & 4.54 & 9.75\tabularnewline
\hline 
\end{tabular} }\caption{\label{tab:parameters}This table shows two important estimates for
the models: the volatility $\sigma$ and the expected frequency $\lambda$
of jumps.}
\end{table}

And finally, we check stability of the estimates from the option data.
First, let us see Figure \ref{fig:implied_density}, which shows the
L\'evy densities $dv/dx$ estimated by deploying the models. The
implied density $dv/dx$ for the ELNN seems to maintain its shape
over time. However, the density values for the Merton and Kou models
tend to rise and fall substantially as time passes. The instabilities
by model misspecification are inevitable, but it must not be severe
because fluctuating estimates bring about totally unreliable predictions.
Furthermore, recall that a slight change of $\Phi_{X_{T}}\left(w-i\right)$
can lead to quite a different $dv/dx$, which is an observed fact
in the previous section. Considering that the fitting errors for $\Phi_{X_{T}}^{*}\left(w-i\right)$
of the Merton and Kou models are much larger than those of the ELNN,
their L\'evy densities are hard to accept. Additionally, we analyze
two important estimates for the models, the volatility $\sigma$ and
the expected frequency $\lambda$ of jumps, through Table \ref{tab:parameters}.
By the figures in the table, in comparison with the parametric models,
the ELNN gives stable values for $\lambda$ but does not for $\sigma$.
In our view, this is because market volatility changes over time in
practice. Thus, it may be supplemented by reflecting stochastic volatility
to the ELNN. 

\section{\label{sec:Conclusion}Conclusion}

In this paper, we propose the exponential L\'evy neural network,
which is a new non-parametric exp-L\'evy model using artificial neural
networks. The ELNN is a strong hybrid model that it fully combines
the ANNs with a conventional pricing model, the exp-L\'evy model.
So, the ELNN can avoid several essential issues of the non-hybrid
models and the weak hybrid models such as unacceptable outcomes and
inconsistent pricing of over-the-counter products. In addition, the
ELNN is the first applicable non-parametric exp-L\'evy model in that
the existing non-parametric models are too vulnerable to be employed
in practice. This can be achieved by virtue of outstanding researches
on optimization in the field of ANN. The empirical tests with S\&P
500 option prices show that the ELNN outperforms two parametric models,
the Merton and Kou models, in terms of fitting performance and stability
of estimates. On the other hand, option prices should be transformed
by the Fourier transform for a learning of the ELNN. In actual markets,
daily data is not enough to be transformed precisely. By inventing
a technique ``data amplification,'' we resolve the problem. Its
effect is sufficiently verified through the tests.

There may be many further topics for the ELNN. Among them, we select
and suggest the followings. First, the ELNN should be extended so
that it can handle infinite L\'evy measures also. Then, the ELNN
would be better than relevant models such as the CGMY model. Moreover,
hidden states such as stochastic volatility should be considered to
accommodate the heteroscedasticity of market returns. Lastly, the
design of the ELNN needs to be improved, for examples, more flexible
activation functions, deeper layers and so on.

\section*{Funding}

This research did not receive any specific grant from funding agencies
in the public, commercial, or not-for-profit sectors.

\bibliographystyle{elsarticle-harv}
\bibliography{ref}

\end{document}